\pdfoutput=1

\documentclass[aps,prl,10pt,twocolumn,showpacs,superscriptaddress,footnoteinbib]{revtex4}
\usepackage{amsmath}
\usepackage{latexsym}
\usepackage{amssymb}
\usepackage{graphics,epstopdf}

\begin{document}
\title{Entanglement witness operator for quantum teleportation}
\author{Nirman Ganguly}
\thanks{nirmanganguly@gmail.com}
\affiliation{Dept. of Mathematics, Heritage Institute of Technology, Kolkata-107, West Bengal, India}
\affiliation{S. N. Bose National Centre for Basic Sciences, Salt Lake,
Kolkata-700 098, India}
\author{Satyabrata Adhikari}
\thanks{tapisatya@iopb.res.in}
\affiliation{Institute of Physics, Sainik School Post, Bhubaneshwar-751005, Orissa, India}
\author{A. S. Majumdar}
\thanks{archan@bose.res.in}
\affiliation{S. N. Bose National Centre for Basic Sciences, Salt Lake,
Kolkata-700 098, India}
\author{Jyotishman Chatterjee}
\thanks{jyotishman.chatterjee@heritageit.edu}
\affiliation{Dept. of Mathematics, Heritage Institute of Technology, Kolkata-107, West Bengal, India}
\date{\today}

\begin{abstract}
The ability of entangled states to act as resource for teleportation 
is linked to a property of the fully entangled fraction. We show that
the set of states with their fully entangled fraction bounded by a threshold
value required for performing teleportation is both convex and compact. 
This feature enables for the
existence of hermitian witness operators the measurement of which could
distinguish unknown states useful for performing teleportation. We present
an example of such a witness operator illustrating it for different classes of
states.   
\end{abstract}


\maketitle

\paragraph{A. Introduction.\textemdash}

Quantum information processing is now widely recognized as a powerful tool
for implementing tasks that cannot be performed using classical means 
\cite{nielsen}. A large number of algorithms for various information
processing tasks such as super dense coding\cite{dense}, 
teleportation \cite{teleport} and key generation 
\cite{crypto}  have been proposed and 
experimentally demonstrated.  At the practical
level information processing is implemented by manipulating states of
quantum particles, and it is well known that not all quantum states can be
used for such purposes. Hence, given an unknown state, one of the most 
relevant issues here is to determine whether it is useful for quantum 
information processing.     

The key ingredient for performing many information processing tasks is 
provided by quantum entanglement. The experimental detection of entanglement is
facilitated by the existence of entanglement witnesses \cite{horodecki,terhal}
which are hermitian operators with at least one negative eigenvalue.  
The existence of entanglement witnesses is a
consequence of the Hahn-Banach theorem in functional 
analysis \cite{hahn1,hahn2} providing a necessary and sufficient condition to 
detect entanglement. Motivated by the nature of different 
classes of entangled states, various methods have been suggested to construct 
entanglement witnesses \cite{lewenstein,terhal2,sperling,adhikari}. 
Study of entanglement witnesses \cite{review} has proceeded in
directions such as the construction of optimal witnesses 
\cite{lewenstein,sperling},
Schmidt number witnesses \cite{sanpera}, and common witnesses
\cite{wu}. The possibility of experimental
detection of entanglement through the
measurement of expectation values of  witness operators for
unknown states is facilitated by the decomposition of witnesses
in terms of Pauli spin matrices for qubits \cite{guhne} and Gell-Mann 
matrices in higher dimensions \cite{bertl}. For macroscopic systems
the properties of thermodynamic quantities provide a useful avenue for detection
of entanglement \cite{vedral}.

Teleportation \cite{teleport} is a typical information processing task
where at present there is intense activity in extending the experimental 
frontiers \cite{telex}. However, it is well known
that not all entangled states are useful for teleportation. For example,
while the entangled Werner state \cite{werner} in $2 \otimes 2$ dimensions is
a useful resource  \cite{lee}, another class of  maximally entangled mixed 
states \cite{mems}, as well as other non-maximally entangled mixed 
states achieve a fidelity higher than the classical limit only
when their magnitude of entanglement exceeds a certain value \cite{adhikari2}. 
The problem of determining states useful for teleportation becomes 
conceptually more involved
in higher dimensions where bound entangled states \cite{bound} also exist.

The motivation for this study is to enquire how to
determine whether an unknown entangled state could be used as a resource 
for performing information processing tasks. In the present work we consider
this question for the specific task of quantum teleportation. We propose 
and demonstrate the existence of 
measurable witness operators connected to teleportation, by 
making use of a property of entangled states, {\it viz},
the fully entangled fraction (FEF) \cite{ben,horodecki2} which 
can be related to the efficacy of teleportation. 
In spite of the conceptual
relevance of the FEF as a characteristic trait of entangled 
states \cite{vidal}, its actual determination could be complicated for
higher dimensional systems\cite{zhao,gu}.  Our proof of the existence
of witnesses connected to a relevant threshold value for the FEF enables us 
to construct a suitable
witness operator for teleportation, as is illustrated with certain examples.

\paragraph{B. Proof of existence of witness.\textemdash}

The fully entangled fraction (FEF) \cite{horodecki2} is defined for a bipartite
state  $\rho$ in $d \otimes d$ dimensions as 
\begin{equation}
F(\rho)= max_{U}\langle \psi^{+} \vert U^{\dagger} \otimes I \rho U \otimes I \vert \psi^{+} \rangle
\label{fef}
\end{equation}
where $\vert \psi^{+} \rangle = \frac{1}{\sqrt{d}} \sum_{i=0}^{d-1} \vert ii \rangle $ and $U$ is a unitary operator.  A quantum channel is useful for 
teleportation if it can 
provide a fidelity higher than what can be done classically. The fidelity 
depends on the FEF of the state, e.g., a state in $d \otimes d$ dimensions 
works as a teleportation channel if its FEF $ > \frac{1}{d}$ 
\cite{horodecki2,vidal,zhao}. 

Here we propose the existence of a hermitian operator which serves 
to distinguish between states having FEF higher than a given threshold value 
from other states.  FEF $>\frac{1}{d}$ is a benchmark which 
measures the viability of quantum states in teleportation. Let 
us consider the set $S$ of states having FEF $\leq \frac{1}{d}$.   
A special geometric form of the Hahn-Banach theorem in functional 
analysis \cite{hahn1,hahn2} states that 
if a set is convex and compact, then a point lying outside the set can be 
separated from it by a hyperplane. The existence of entanglement witnesses
are indeed also an outcome of this theorem \cite{horodecki,terhal}.
We now present the proof that the set $S$ of states with FEF $\leq \frac{1}{d}$ 
is indeed convex and compact, so that the separation axiom in the form of the 
Hahn-Banach theorem could be applied in order to demonstrate the existence of 
hermitian witness operators  for 
teleportation. 

\noindent \textit{Proposition:} The set $S=\lbrace \rho: F(\rho) \leq \frac{1}{d} \rbrace$ is convex and compact.
\textit{Proof:} The proof is done in two steps.
\textit{(i)  We  first show that $S$ is convex}. Let $\rho_{1},\rho_{2} \in S$. Therefore, 
\begin{equation}
F(\rho_{1})\leq \frac{1}{d}, ~~~ F(\rho_{2})\leq \frac{1}{d}.
\label{frho12}
\end{equation}
Consider $\rho_{c}=\lambda \rho_{1} + (1-\lambda) \rho_{2}$, where $\lambda \in [0,1]$ and $F(\rho_{c})= \langle \psi^{+} \vert U_{c}^{\dagger} \otimes I \rho_{c} U_{c} \otimes I \vert \psi^{+} \rangle$. Now,
$F(\rho_{c})= \lambda \langle \psi^{+} \vert U_{c}^{\dagger} \otimes I \rho_{1} U_{c} \otimes I \vert \psi^{+} \rangle 
+(1-\lambda)\langle \psi^{+} \vert U_{c}^{\dagger} \otimes I \rho_{2} U_{c} \otimes I \vert \psi^{+} \rangle$. Let $F(\rho_{i})= \langle \psi^{+} \vert U_{i}^{\dagger} \otimes I \rho_{i} U_{i} \otimes I \vert \psi^{+} \rangle,~~(i=1,2)$. This is 
possible since the group of 
unitary matrices is compact, hence the maximum will be attained for a unitary 
matrix $U$. It follows that
$F(\rho_{c}) \leq \lambda F(\rho_{1})+ (1-\lambda)F(\rho_{2}).$ 
Using Eq.(\ref{frho12}) we have
\begin{eqnarray}
F(\rho_{c}) \leq \frac{1}{d}
\label{convex}
\end{eqnarray}
Thus, $\rho_{c}$ lies in $S$, and hence, $S$ is convex.

\textit{(ii) We now show that $S$ is compact}. 
Note that in a finite dimensional Hilbert space, in order to 
show that a set is compact it is enough to show that the set is closed and 
bounded. The set $S$ is bounded as every density matrix has a bounded spectrum, 
i.e., eigenvalues lying between $0$ and $1$. In order to prove that
 the set $S$ is closed, consider first  the following lemma.
\textit{Lemma}: Let $A$ and $B$ be two matrices of size $m \times n$ and $n \times r$ respectively. Then $ \Vert AB \Vert \leq \Vert A \Vert \Vert B \Vert $,
where the norm of a matrix $A$  is defined as 
$\Vert A \Vert = \sqrt{TrA^{\dagger}A}=\sqrt{\sum_{i}\sum_{j}\vert A_{ij}  \vert^{2}}$.\\
\textit{Proof of the lemma}: Let $A= \left(%
\begin{array}{c}
  A_{1} \\
  A_{2}  \\
  .\\
  .\\
  A_{m}
\end{array}%
\right)$
 and $B=[B^{(1)} B^{(2)}.... B^{(r)}]$ , where $A_{i}$'s are row vectors of size $n$ and $B^{(j)}$'s are column vectors of size $n$ respectively. Using the Cauchy-Schwarz 
inequality, it follows that
$\vert (AB)_{ij} \vert = \vert A_{i}B^{(j)} \vert \leq \Vert A_{i} \Vert \Vert B^{(j)} \Vert.$
Therefore, one has 
\begin{eqnarray}
\Vert AB \Vert^{2} = \sum_{i=1}^{m}\sum_{j=1}^{r}\vert (AB)_{ij} \vert ^{2} 
\leq  \sum_{i=1}^{m}\sum_{j=1}^{r} \Vert A_{i} \Vert^{2} \Vert B^{(j)} \Vert^{2} 
\label{lemma}
\end{eqnarray}
The r.h.s of the above inequality can be expressed as 
$\sum_{i=1}^{m} \Vert A_{i} \Vert^{2} \sum_{j=1}^{r} \Vert B^{(j)} \Vert^{2} 
= \Vert A \Vert^{2} \Vert B \Vert^{2}$, from which it follows that
$ \Vert AB \Vert \leq \Vert A \Vert \Vert B \Vert $.

For any two density matrices $\rho_{a}$ and $\rho_{b}$, assume the 
maximum value of FEF is obtained at $U_{a}$ and $U_{b}$ respectively, i.e.,
 $F(\rho_{a})= \langle \psi^{+} \vert U_{a}^{\dagger} \otimes I \rho_{a} U_{a} \otimes I \vert \psi^{+} \rangle$ and $F(\rho_{b})= \langle \psi^{+} \vert U_{b}^{\dagger} \otimes I \rho_{b} U_{b} \otimes I \vert \psi^{+} \rangle$. Therefore, we have
$F(\rho_{a})-F(\rho_{b})=
\langle \psi^{+} \vert U_{a}^{\dagger} \otimes I \rho_{a} U_{a} \otimes I \vert \psi^{+} \rangle - \langle \psi^{+} \vert U_{b}^{\dagger} \otimes I \rho_{b} U_{b} \otimes I \vert \psi^{+} \rangle$ from which it follows that $F(\rho_{a})-F(\rho_{b})
\leq \langle \psi^{+} \vert U_{a}^{\dagger} \otimes I \rho_{a} U_{a} \otimes I \vert \psi^{+} \rangle - \langle \psi^{+} \vert U_{a}^{\dagger} \otimes I \rho_{b} U_{a} \otimes I \vert \psi^{+} \rangle$ since $\langle \psi^{+} \vert U_{a}^{\dagger} \otimes I \rho_{b} U_{a} \otimes I \vert \psi^{+} \rangle \leq \langle \psi^{+} \vert U_{b}^{\dagger} \otimes I \rho_{b} U_{b} \otimes I \vert \psi^{+} \rangle$. Hence,
$F(\rho_{a})-F(\rho_{b}) \leq
\langle \psi^{+} \vert U_{a}^{\dagger} \otimes I (\rho_{a}-\rho_{b}) U_{a} \otimes I \vert \psi^{+} \rangle$, implying 
\begin{equation}
F(\rho_{a})-F(\rho_{b})
\leq \vert \langle \psi^{+} \vert U_{a}^{\dagger} \otimes I (\rho_{a}-\rho_{b}) U_{a} \otimes I \vert \psi^{+} \rangle \vert. 
\end{equation}
Now, using the above lemma, one gets
$F(\rho_{a})-F(\rho_{b}) \leq \Vert \langle \psi^{+} \vert \Vert \Vert U_{a}^{\dagger} \otimes I \Vert \Vert (\rho_{a}-\rho_{b}) \Vert \Vert U_{a} \otimes I \Vert \Vert \vert \psi^{+} \rangle \Vert$, or $F(\rho_{a})-F(\rho_{b})
\leq C^{2}K_{1}^{2} \Vert \rho_{a}-\rho_{b}  \Vert $, where $C, K_1$
are positive real numbers. The last step follows from the fact that 
$\Vert \langle \psi^{+} \vert \Vert = C$. Since the set of all unitary 
operators is compact, it is bounded,  and thus for any $U$, $\Vert U \otimes I \Vert \leq K_{1}$. Similarly $F(\rho_{b})-F(\rho_{a})
\leq C^{2}K_{1}^{2} \Vert \rho_{b}-\rho_{a}  \Vert = C^{2}K_{1}^{2} \Vert \rho_{a}-\rho_{b}  \Vert $. So finally, one may write
\begin{equation}
\vert F(\rho_{a})-F(\rho_{b}) \vert \leq
 C^2K_1^2\Vert \rho_{a}-\rho_{b} \Vert. 
\label{compact}
\end{equation}
This implies that $F$ is a continuous function. 
Moreover, for any density matrix $\rho$, with 
$F(\rho) \in [\frac{1}{d^{2}},1]$, 
one has $F(\rho)=1$ iff $\rho$ is a maximally entangled pure state, 
and $F(\rho)= \frac{1}{d^{2}}$ iff $\rho$ is the maximally mixed 
state \cite{zhao}. For the set $S$ in our consideration $F(\rho) \in [\frac{1}{d^{2}},\frac{1}{d}]$.
Hence, $S=\lbrace \rho: F(\rho) \leq \frac{1}{d} \rbrace = F^{-1}([\frac{1}{d^{2}},\frac{1}{d}])$, is closed \cite{hahn2}. This completes the proof 
of our proposition
that the set $S=\lbrace \rho: F(\rho) \leq \frac{1}{d} \rbrace$ is convex and compact.

It now follows from the Hahn-Banach theorem \cite{hahn1,hahn2}, that any $\chi \not \in S$ can be separated from $S$ by a hyperplane. 
In other words, any state useful for teleportation can be separated from the states not useful for teleportation by a hyperplane and thus allows for the definition of a witness. The witness operator, if so defined, identifies the states which 
are useful in the teleportation protocol, i.e., provides a fidelity higher than 
the classical optimum. 

\begin{figure}[!ht]
\resizebox{7cm}{7cm}{\includegraphics{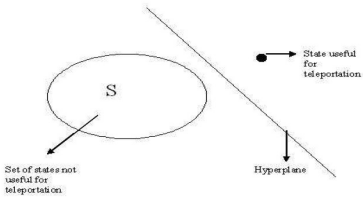}}
\vskip -1.0in
\caption{The set $S=\lbrace \rho: F(\rho) \leq \frac{1}{d} \rbrace$ is convex and 
compact, and using the
Hahn-Banach theorem it follows that any state useful for teleportation can be separated from the states not useful for teleportation by a hyperplane, 
thus providing for the existence of a witness for teleportation.}
\end{figure}

\paragraph{C. A witness operator for teleportation.\textemdash}

A hermitian operator $W$ may be called a teleportation witness if the 
following conditions are satisfied: 
(i) $Tr(W\sigma)\geq 0$, for all states $\sigma$ which are not useful for 
teleportation.
(ii) $Tr(W\chi) < 0$, for at least one state $\chi$ which is useful for 
teleportation.
We propose a hermitian operator for a $d \otimes d$ system of the form
(using $\vert \psi^{+} \rangle = \frac{1}{\sqrt{d}}\sum_{i=0}^{d-1}\vert ii \rangle$)
\begin{equation}
W = \frac{1}{d}I - \vert \psi^{+} \rangle \langle \psi^{+} \vert 
\label{TW}
\end{equation}
In order to prove that $W$  is indeed a witness operator, we first  show that
the operator $W$ gives a non-negative expectation over all states which are 
not useful for teleportation. Let $\sigma$ be an arbitrary state chosen from 
the set $S$ not useful for teleportation, i.e., $\sigma \in S$.  Hence,
\begin{equation}
Tr(W\sigma)= \frac{1}{d} - \langle \psi^{+} \vert \sigma \vert \psi^{+} \rangle
\end{equation}
from which it follows that
$Tr(W\sigma) \geq \frac{1}{d} - max_{U}\langle \psi^{+} \vert U^{\dagger} \otimes I \sigma U \otimes I \vert \psi^{+} \rangle$.
Now, using the definition of the FEF, $F(\sigma)$ from Eq.(\ref{fef}), and the 
fact that  $\sigma \in S$, one gets
\begin{equation}
Tr(W\sigma) \geq 0
\end{equation}
Our task now is to show that the operator $W$ detects at least one entangled 
state $\chi$ which is useful for teleportation,  i.e.,  $Tr(W\chi)<0$, which
we do by providing the following illustrations.

Let us first consider the isotropic state 
\begin{equation}
\chi_{\beta}= \beta \vert \psi^{+} \rangle \langle \psi^{+} \vert + \frac{1-\beta}{d^{2}}I~~~~~~~~~(-\frac{1}{d^{2}-1}\leq \beta \leq 1)
\end{equation}
The isotropic state is entangled $\forall \beta > \frac{1}{d+1}$ \cite{bertl2}. 
Now,
$Tr(W\chi_{\beta})=\frac{(d-1)(1-\beta(d+1))}{d^{2}}$,
from which it follows that $Tr(W\chi_{\beta})< 0$, when $\beta>\frac{1}{d+1}$. 
Therefore, all entangled isotropic states are useful for teleportation. The 
same conclusion was obtained in  Ref. \cite{zhao}  on explicit 
calculation of the FEF for isotropic states. 
We next consider the generalized Werner state \cite{werner,pitt} in 
$d \otimes d$ given by
\begin{equation}
\chi_{wer}=(1-v)\frac{I}{d^{2}}+ v\vert \psi_{d}\rangle \langle \psi_{d} \vert
\end{equation}
where $0 \leq v \leq 1$ and $\vert \psi_{d}\rangle = \sum_{i=0}^{d-1}\alpha_{i}\vert ii \rangle$, with   $\sum_{i}\vert \alpha_{i} \vert^2=1$,
for which one obtains
$Tr(W\chi_{wer})=\frac{1}{d}-\frac{1-v}{d^{2}}-\frac{v}{d}\sum_{i=0}^{d-1}\alpha_{i}\sum_{i=0}^{d-1}\alpha_{i}^{*}$.
The witness $W$ detects those Werner states which are useful for teleportation,
i.e.,  $Tr(W\rho_{wer})< 0$, which is the case when
\begin{equation}
\frac{1}{d}-\frac{1-v}{d^{2}}-\frac{v}{d}\sum_{i=0}^{d-1}\alpha_{i}\sum_{i=0}^{d-1}\alpha_{i}^{*} < 0
\end{equation}
In $2 \otimes 2$ dimensions with $\alpha_i = 1/\sqrt{2}$, one gets
$Tr(W\chi_{wer})=\frac{1-3v}{4}<0, \text{when}~~ v>\frac{1}{3}$. Thus, all
entangled Werner states are useful for teleportation, a result which is
well-known \cite{lee}.

Now, consider another class of maximally entangled mixed states in 
$2 \otimes 2$ dimensions, which possess
the maximum amount of entanglement for a given purity \cite{mems}:
\begin{equation}
\chi_{MEMS}= \left(%
\begin{array}{cccc}
  h(C) & 0 & 0 & C/2 \\
  0 & 1-2h(C) & 0 & 0  \\
  0 & 0 & 0 & 0  \\
  C/2 & 0 & 0 & h(C)  \\
\end{array}%
\right)
\label{mems}
\end{equation}
where, $h(C)= C/2$ for  $C\geq2/3$, and 
$h(C)= 1/3$ for  $C<2/3$, with 
 $C$ the concurrence of $\chi_{MEMS}$.
Here we obtain
$Tr(W\rho_{MEMS})= \frac{1}{2}-h(C)-\frac{C}{2}$.
It follows that $Tr(W\rho_{MEMS})\geq 0$ when  $0\leq C \leq \frac{1}{3}$,
implying that for a magnitude of the entanglement in the above range, the
state $\chi_{MEMS}$ is not useful for teleportation. But,
for $C>\frac{1}{3}$, the state $\chi_{MEMS}$ is suitable for teleportation, as
one obtains  $Tr(W\rho_{MEMS}) < 0$ in this case, confirming the results 
derived earlier in the literature \cite{adhikari2}.
However, as expected with any witness, our
proposed witness operator may fail to identify  certain other 
states that are known
to be useful for teleportation. For example, the state
(for $\vert \phi \rangle = \frac{1}{\sqrt{2}}(\vert 01 \rangle + \vert 10 \rangle)$ and  $0 \leq a \leq 1$)
\begin{equation}
\rho_{\phi} = a\vert \phi \rangle \langle \phi \vert + (1-a)\vert 11 \rangle \langle 11 \vert 
\end{equation}
was recently studied in the context of quantum
discord \cite{ali}. This class of states is useful for teleportation but the
 witness $W$ is unable to detect it as $Tr(W\rho_{\phi})=\frac{a}{2}\geq 0$.

Let us now briefly discuss 
the measurability of the witness operator. 
For experimental realization  of the witness it is
necessary to decompose the witness into operators that can be measured locally,
 i.e, a decomposition into projectors of the form 
$W = \sum_{i=1}^{k}c_{i}\vert e_{i}\rangle \langle e_{i}\vert \otimes \vert f_{i}\rangle \langle f_{i}\vert$ \cite{review,guhne}.
For implementation using polarized photons as in \cite{barbieri}, one may
 take $\vert H \rangle = \vert 0 \rangle ,\vert V \rangle = \vert 1 \rangle , \vert D \rangle = \frac{\vert H \rangle + \vert V \rangle }{\sqrt{2}}, \vert F \rangle = \frac{\vert H \rangle - \vert V \rangle }{\sqrt{2}}, \vert L \rangle = \frac{\vert H \rangle + i\vert  V \rangle }{\sqrt{2}}, \vert R \rangle = \frac{\vert H \rangle - i\vert  V \rangle }{\sqrt{2}} $ as the horizontal, vertical, 
diagonal, and the left and right circular polarization states, respectively.
Using a result given in \cite{hyllus},  our witness operator 
 can be recast for qubits into the required form, given by
\begin{eqnarray}
W = \frac{1}{2}(\vert HV \rangle \langle HV \vert + \vert VH \rangle \langle VH \vert - \vert DD \rangle \langle DD \vert \nonumber \\
- \vert FF \rangle \langle FF \vert +\vert LL \rangle \langle LL \vert + \vert RR \rangle \langle RR \vert)
\end{eqnarray}
Using this technique
for an unknown two-qubit state $\chi$, the estimation of 
$\langle W \rangle$  requires three measurements \cite{hyllus}, as is also 
evident from the decomposition 
of our witness operator for qubits in terms of Pauli 
spin matrices, i.e., 
$W=\frac{1}{4}[I \otimes I - \sigma_{x} \otimes \sigma_{x} + \sigma_{y} \otimes \sigma_{y} - \sigma_{z} \otimes \sigma_{z}]$,
which is far less than the measurement of $15$ parameters 
required for full state 
tomography\cite{munro}. In higher dimensions, the witness operator may be
decomposed in terms of Gell-Mann matrices \cite{bertl}, and this difference 
further increases with the
increase in dimensions. Therefore, the utility of the  witness operator is
indicated as compared to full state tomography when discrimination 
of useful entangled states for performing teleportation 
is required.

Before concluding, it may be noted that is possible to relate the FEF
(\ref{fef}) with the maximum fidelity for other information processing tasks, 
such as super dense coding and entanglement swapping  \cite{grond}.
In the generalized dense coding for $d \otimes d$ systems, one can use a 
maximally  entangled state $\vert\phi\rangle$ to encode $d^{2}/2$ bits in 
$d^{2}$ orthogonal states $(I \otimes U_{i})\vert\phi\rangle$ \cite{liu}. 
If the maximally 
entangled state is replaced with a general density operator, the dense 
coding fidelity is defined as an average over the $d^{2}$ results. 
A relation between the maximum fidelity $F_{DC}^{max} $ of dense coding and 
the FEF was established for $d \otimes d$ systems to be
$F_{DC}^{max} = F$. Similarly,
for two-qubit systems the maximum fidelity of entanglement swapping\cite{swap}
$F_{ES}^{max}$ is also related to the FEF by
$F_{ES}^{max} = F$  \cite{grond}. However, teleportation is a different
information processing task as compared to dense coding where $F > 1/d$
does not guarantee a higher than classical fidelity \cite{bru}. Hence, it
is not possible to apply the  above witness (\ref{TW}) to super dense 
coding and entanglement swapping. 

\paragraph{D. Conclusions.\textemdash}

To summarize, in this work we have proposed a framework for discriminating
quantum states useful for performing teleportation
through the measurement of a hermitian witness operator. The ability
 of an entangled state to act as a resource for 
teleportation is connected with the fully entangled fraction of the state.
The estimation of the fully entangled fraction is difficult in general,
except in the case of some known states. We have shown that the
set of states having their fully entangled fraction bounded by a certain
threshold value required for teleportation
is both convex and compact. Exploiting this feature we have
demonstrated the existence of a witness operator for teleportation.
The measurement of the expectation value of the witness for unknown states
reveals which states are useful as resource for performing
teleportation. We have provided some illustrations of the applicability of
the witness for isotropic and 
Werner states in $d \otimes d$ dimensions, and another class of maximally
entangled mixed states for qubits.
The measurability of such a witness operator requires determination of a much
lesser number of parameters in comparison to state tomography of an unknown
state, thus signifying the practical utility of our proposal. It would be 
interesting to explore the possibility of existence of witnesses for various 
other information processing tasks, as well. In this context further studies
on finding optimal and common witnesses are called for.

\emph{Acknowledgments:} ASM would like to acknowledge support from the
DST Project SR/S2/PU-16/2007.


\begin{thebibliography}{99}

\bibitem{nielsen} M. A. Nielsen and I. L. Chuang, {\it Quantum Computation 
and Quantum Information}, (Cambridge University Press, Cambridge, 2000).

\bibitem{dense} C. H. Bennett and S. J. Wiesner, Phys. Rev. Lett. {\bf 69},
 2881  (1992).

\bibitem{teleport} C. H. Bennett, G. Brassard, C. Crepeau, R. Jozsa, A. Peres,
and W. K. Wootters, Phys. Rev. Lett. {\bf 70},  1895  (1993).

\bibitem{crypto} A. K. Ekert, Phys. Rev. Lett. {\bf 67},  661 (1991).

\bibitem{horodecki} M. Horodecki, P. Horodecki, and R. Horodecki, Phys. Lett. 
A {\bf 223}, 1  (1996).

\bibitem{terhal} B. M. Terhal, Phys. Lett. A {\bf 271},  319 (2000).

\bibitem{hahn1}	R. B. Holmes, {\it Geometric Functional Analysis and its Applications}, (Springer Verlag, 1975).

\bibitem{hahn2} W. Rudin, {\it Principles of Mathematical Analysis}, (McGraw Hill, 1976).

\bibitem{lewenstein} M. Lewenstein, B. Krauss, J. I. Cirac and P. Horodecki, 
Phys. Rev. A {\bf 62},  052310  (2000).

\bibitem{terhal2} B. M. Terhal, J. Theor. Comput. Sci. {\bf 287},  313 (2002).

\bibitem{sperling} J. Sperling and W. Vogel, Phys. Rev. A {\bf 79},  022318 
 (2009).

\bibitem{adhikari} N. Ganguly and S. Adhikari, Phys. Rev. A {\bf 80}, 032331
(2009).

\bibitem{review} O. Guhne and  G. Toth, Phys. Rep. {\bf 474}, 1  (2009). 

\bibitem{sanpera} B. M. Terhal and P. Horodecki, Phys. Rev. A {\bf 61}, 040301R 
(2000); A. Sanpera, D. Bru$\beta$, M. Lewenstein, Phys. Rev. A 
{\bf 63}, 050301R (2001).

\bibitem{wu} Y. C. Wu and G. C. Guo, Phys. Rev. A{\bf 75}, 052333 (2007);
N. Ganguly, S. Adhikari and A. S. Majumdar, arXiv: 1101.0477.

\bibitem{guhne}	O. Guhne, P. Hyllus, D. Bru$\beta$, A. Ekert, M. Lewenstein, 
C. Macchiavello and A. Sanpera, Phys. Rev. A {\bf 66}, (2002) 062305.

\bibitem{bertl}	R. A Bertlmann and P. Krammer, J. Phys. A: Math. Theor. 
{\bf 41},  235303 (2008).

\bibitem{vedral} C. Brukner and V. Vedral, arXiv:quant-ph/0406040; 
G. Toth, Phys. Rev. A {\bf 71}, 010301(R) (2005); J. Hide, W. Son and V. 
Vedral, Phys. Rev. Lett. {\bf 102}, 100503 (2009).

\bibitem{telex} D. Bouwmeester, J.-W. Pan, K. Mattle, M. Eibl, H. Weinfurter
and A. Zeilinger, Nature {\bf 390}, 575 (1997); S. Olmschenk, D. N. 
Matsukevich, P. Maunz, D. Hayes, L.-M. Duan and C. Monroe, Science {\bf 323}, 
486 (2009); X.-M. Jin et al., Nature Photonics {\bf 4}, 376 (2010).

\bibitem{werner} R. F. Werner, Phys. Rev. A {\bf 40},  4277 (1989).

\bibitem{lee} J. Lee and M. S. Kim, Phys. Rev. Lett. {\bf 84}, 4236 (2000).

\bibitem{mems}  W. J. Munro, D. F. V. James, A. G. White and P. G. Kwiat,
Phys. Rev. A {\bf 64},  030302 (2001).

\bibitem{adhikari2} S. Adhikari, A. S. Majumdar, S. Roy, B. Ghosh and N. Nayak,
Quant. Inf. Comm. {\bf 10}, 0398 (2010). 

\bibitem{bound}  P. Horodecki, Phys. Lett. A {\bf 232}, 333 (1997);
M. Horodecki, P. Horodecki, and R. Horodecki, Phys. Rev. Lett. {\bf 80}, 5239 
(1998).

\bibitem{ben} C. H. Bennett, D. P. DiVincenzo, J. A. Smolin and W. K. Wootters,
Phys. Rev. A {\bf 54}, 3824 (1996).

\bibitem{horodecki2} M. Horodecki, P. Horodecki and R. Horodecki, Phys. Rev. 
A {\bf 60}, 1888  (1999).

\bibitem{vidal} G. Vidal, D. Jonathan and M. A. Nielsen, Phys. Rev. A {\bf 62},
 012304 (2000).

\bibitem{zhao} M. J. Zhao, Z. G. Li, S. M. Fei and Z. X. Wang, J. Phys. A: 
Math. Theor. {\bf 43},  275203 (2010).

\bibitem{gu}  R. J. Gu, M. Li, S. M. Fei and X. Q. 
Li-Jost, Commun. Theor. Phys. {\bf 53},   265  (2010).

\bibitem{swap} M. Zukowski, A. Zeillinger, M. A. Horne and A. K. Ekert, Phys. 
Rev. Lett {\bf 71},  4287 (1993).

\bibitem{grond} J. Grondalski, D. M. Etlinger and D. F. V. James, Phys. Lett. A
 {\bf 300},  573   (2002).

\bibitem{ali} M. Ali, A. R. P. Rau and G. Alber, Phys. Rev. A {\bf 81}, 042105
 (2010).

\bibitem{liu} X. S. Liu, G. L. Long, D. M. Tong and F. Li, Phys. Rev. 
A {\bf 65},  022304 (2002).

\bibitem{bertl2} R. A. Bertlmann, K. Durstberger, B. C. Hiesmayr and P.Krammer,
 Phys. Rev. A {\bf 72},  052331 (2005).

\bibitem{pitt} A. O. Pittenger and M. H. Rubin, Opt. Comm. {\bf 179},  447
 (2000);  D. L. Deng and J. L. Chen, Ann. Phys. {\bf 324}, 408   (2009).

\bibitem{barbieri} M. Barbieri, F. De Martini, G. Di Nepi, P. Mataloni, G. M.
 D'Adriano and C. Macchiavello, Phys. Rev. Lett \textbf{91}, 227901  (2003).

\bibitem{hyllus} P.Hyllus, ``Witnessing entanglement in qudit systems'', PhD 
Thesis, University of Hannover (2005).

\bibitem{munro} D. F. V. James, P. G. Kwiat, W. J. Munro and A. G. White,
Phys. Rev. A {\bf 64},  052312 (2001).

\bibitem{bru} D. Bru$\beta$, G. M. D'Ariano, M. Lewenstein, C. Macchiavello,
A. Sen De and U. Sen, Phys. Rev. Lett. {\bf 93}, 210501 (2004).


\end{thebibliography}
\end{document}